\documentclass[12pt]{article}

\usepackage{epsf}

\usepackage{rotate}

\renewcommand{\thefootnote}{\fnsymbol{footnote}}

\begin{document}


\begin{titlepage}
\begin{flushright}
\begin{tabular}{l}
DESY 00--014\\
hep--ph/0001253\\
January 2000
\end{tabular}
\end{flushright}

\vspace*{2truecm}

\begin{center}
\boldmath
{\Large \bf Constraining Penguin Contributions and the\\

\vspace*{0.3truecm} 

CKM Angle $\gamma$ through $B_d\to\pi^+\pi^-$}

\unboldmath

\vspace*{2.5cm}

{\sc{\large Robert Fleischer}}\footnote{E-mail: 
{\tt Robert.Fleischer@desy.de}}

\vspace*{0.4cm} 

{\it Deutsches Elektronen-Synchrotron DESY, 
Notkestr.\ 85, D--22607 Hamburg, Germany}

\vspace{3truecm}

{\large\bf Abstract\\[10pt]} \parbox[t]{\textwidth}{
The decays $B_d\to\pi^+\pi^-$ and $B_s\to K^+K^-$ provide an interesting 
strategy to extract the CKM angle $\gamma$ at ``second-generation'' 
$B$-physics experiments of the LHC era. A variant for ``first-generation'' 
experiments can be obtained, if $B_s\to K^+K^-$ is replaced by 
$B_d\to\pi^\mp K^\pm$. We show that the most recent experimental results 
for the CP-averaged $B_d\to\pi^+\pi^-$ and $B_d\to\pi^\mp K^\pm$ branching 
ratios imply a rather restricted range for the corresponding penguin
parameters, and upper bounds on the direct CP asymmetries
${\cal A}_{\rm CP}^{\rm dir}(B_d\to\pi^+\pi^-)$ and 
${\cal A}_{\rm CP}^{\rm dir}(B_d\to\pi^\mp K^\pm)$. Moreover,
we point out that interesting constraints on $\gamma$ can be obtained from 
the CP-averaged $B_d\to\pi^+\pi^-$ and $B_d\to\pi^\mp K^\pm$ branching 
ratios, if in addition mixing-induced CP violation in the former decay
is measured, and the $B^0_d$--$\overline{B^0_d}$ mixing phase is fixed
through $B_d\to J/\psi K_{\rm S}$. An extraction of $\gamma$ becomes 
possible, if furthermore direct CP violation in $B_d\to\pi^+\pi^-$ or 
$B_d\to\pi^\mp K^\pm$ is observed.}

  \vskip1.5cm

\end{center}

\end{titlepage}

\thispagestyle{empty}
\vbox{}
\newpage
 
\setcounter{page}{1}

\setcounter{footnote}{0}
\renewcommand{\thefootnote}{\arabic{footnote}}

\section{Introduction}\label{intro}
Among the central targets of the $B$-factories is a measurement of the 
time-dependent CP asymmetry of the decay $B_d\to\pi^+\pi^-$ \cite{revs}, 
which can be expressed as follows:
\begin{eqnarray}
\lefteqn{a_{\rm CP}(B_d(t)\to\pi^+\pi^-)\equiv
\frac{\mbox{BR}(B_d^0(t)\to \pi^+\pi^-)-
\mbox{BR}(\overline{B_d^0}(t)\to \pi^+\pi^-)}{\mbox{BR}(B_d^0(t)\to 
\pi^+\pi^-)+\mbox{BR}(\overline{B_d^0}(t)\to \pi^+\pi^-)}}\nonumber\\
&&={\cal A}_{\rm CP}^{\rm dir}(B_d\to \pi^+\pi^-)\cos(\Delta M_d t)+
{\cal A}_{\rm CP}^{\rm mix}(B_d\to \pi^+\pi^-)\sin(\Delta M_d t).
\end{eqnarray}
Here ${\cal A}_{\rm CP}^{\rm dir}(B_d\to \pi^+\pi^-)$ and
${\cal A}_{\rm CP}^{\rm mix}(B_d\to \pi^+\pi^-)$ are due to ``direct'' and
``mixing-induced'' CP violation, respectively. In the summer of 1999, 
the CLEO collaboration reported the first observation of the long-awaited 
$B_d\to\pi^+\pi^-$ transition, with the following CP-averaged branching 
ratio \cite{CLEO1}:
\begin{equation}\label{CLEO-res1}
\mbox{BR}(B_d\to\pi^+\pi^-)\equiv\frac{1}{2}\left[
\mbox{BR}(B_d^0\to\pi^+\pi^-)+\mbox{BR}(\overline{B_d^0}\to\pi^+\pi^-)\right]
=\left(4.3^{+1.6}_{-1.4}\pm0.5\right)\times10^{-6}.
\end{equation}
This channel usually appears in the literature as a tool to determine the 
angle $\alpha=180^\circ-\beta-\gamma$ of the unitarity triangle \cite{ut} 
of the Cabibbo--Kobayashi--Maskawa matrix (CKM matrix) \cite{ckm}. 
However, penguin topologies are expected to affect this determination 
severely. Although there are several strategies on the market to control
these penguin uncertainties \cite{revs}, they are usually very challenging 
from an experimental point of view. Constraints on $\alpha$ from 
$B_d\to\pi^+\pi^-$ were considered in \cite{gq}--\cite{pirjol}.

In a recent paper \cite{RF-BsKK}, a strategy was proposed, where 
$B_d\to\pi^+\pi^-$ is combined with its $U$-spin counterpart $B_s\to
K^+K^-$ \cite{Dunietz} to extract $\phi_d=2\beta$ and $\gamma$. If the
phase-convention independent quantity $\phi_d$, which is related to the 
$B^0_d$--$\overline{B^0_d}$ mixing phase and can be determined 
straightforwardly with the help of the ``gold-plated'' mode 
$B_d\to J/\psi K_{\rm S}$ \cite{bisa}, is used as an input, the $U$-spin
arguments in the extraction of $\gamma$ can be minimized. This approach,
which relies only on the $U$-spin flavour symmetry and is {\it not} affected
by any final-state-interaction effects \cite{FSI}, is very promising for 
``second-generation'' $B$-physics experiments at hadron machines, such as LHCb
or BTeV \cite{LHC}. There is a variant of this strategy for the asymmetric 
$e^+e^-$ $B$-factories operating at the $\Upsilon(4S)$ resonance (BaBar and 
BELLE), where $B_s$ decays cannot be explored, if $B_s\to K^+K^-$ is replaced 
by $B_d\to\pi^\mp K^\pm$, and a certain dynamical assumption concerning
``exchange'' and ``penguin annihilation'' topologies is made. Although 
$B_s\to K^+K^-$ should be accessible at HERA-B and Run II of the Tevatron, 
a measurement of $B_d\to\pi^\mp K^\pm$ may be easier for these 
``first-generation'' hadronic $B$ experiments. At \mbox{HERA-B}, for
instance, one expects to collect 260 and 35 decay events per year of 
$B_d\to\pi^\mp K^\pm$ and $B_s\to K^+K^-$, respectively \cite{HERA-B}. 
The present result for the CP-averaged $B_d\to\pi^\mp K^\pm$
branching ratio from the CLEO collaboration is as follows \cite{CLEO1}:
\begin{equation}\label{CLEO-res2}
\mbox{BR}(B_d\to\pi^\mp K^\pm)\equiv\frac{1}{2}\left[
\mbox{BR}(B_d^0\to\pi^-K^+)+\mbox{BR}(\overline{B_d^0}\to\pi^+K^-)\right]
=\left(17.2^{+2.5}_{-2.4}\pm1.2\right)\times10^{-6};
\end{equation}
a first result for the corresponding direct CP asymmetry is also available 
\cite{CLEO2}:
\begin{equation}\label{CLEO-res3}
{\cal A}_{\rm CP}^{\rm dir}(B_d\to\pi^\mp K^\pm)\equiv
\frac{\mbox{BR}(B_d^0\to\pi^-K^+)-
\mbox{BR}(\overline{B_d^0}\to\pi^+K^-)}{\mbox{BR}(B_d^0\to\pi^-K^+)+
\mbox{BR}(\overline{B_d^0}\to\pi^+K^-)}=0.04\pm0.16\,.
\end{equation}

In this paper, we point out that the CLEO results (\ref{CLEO-res1}) and 
(\ref{CLEO-res2}) imply -- among other things -- a rather restricted range 
for the ratio of the ``penguin'' to ``tree'' contributions of the decay 
$B_d\to\pi^+\pi^-$, and upper bounds on the direct CP asymmetries 
${\cal A}_{\rm CP}^{\rm dir}(B_d\to\pi^+\pi^-)$ and 
${\cal A}_{\rm CP}^{\rm dir}(B_d\to\pi^\mp K^\pm)$. If in addition 
mixing-induced CP violation in $B_d\to\pi^+\pi^-$ is measured and $\phi_d$ 
is fixed through $B_d\to J/\psi K_{\rm S}$, we may obtain moreover 
interesting constraints on $\gamma$. An extraction of this angle becomes 
possible, if direct CP violation in $B_d\to\pi^+\pi^-$ or 
$B_d\to\pi^\mp K^\pm$ is observed. 

The outline of this paper is as follows: in Section~\ref{sec:ampl}, 
we have a brief look at the general structure of the relevant decay 
amplitudes and observables. The constraints on the penguin parameters and 
the direct CP asymmetries are discussed in Section~\ref{sec:d-A-const}, 
whereas the bounds on $\gamma$ are the subject of 
Section~\ref{sec:gamma-const}. Finally, the conclusions and an outlook 
are given in Section~\ref{sec:concl}.

\section{Decay Amplitudes and Observables}\label{sec:ampl}
The transition amplitude of the $\bar b\to\bar d$ decay $B_d^0\to\pi^+\pi^-$ 
can be written as follows \cite{RF-rev}:
\begin{equation}\label{Bd-ampl1}
A(B_d^0\to\pi^+\pi^-)=\lambda_u^{(d)}\left(A_{\rm cc}^{u}+A_{\rm pen}^{u}
\right)+\lambda_c^{(d)}A_{\rm pen}^{c}+\lambda_t^{(d)}A_{\rm pen}^{t}\,,
\end{equation}
where $A_{\rm cc}^{u}$ is due to ``current--current'' contributions, the 
amplitudes $A_{\rm pen}^{j}$ describe ``penguin'' topologies with internal 
$j$ quarks ($j\in\{u,c,t\})$, and the 
\begin{equation}
\lambda_j^{(d)}\equiv V_{jd}V_{jb}^\ast
\end{equation}
are the usual CKM factors. Making use of the unitarity of the CKM matrix 
and applying the Wolfenstein parametrization \cite{wolf}, generalized to 
include non-leading terms in $\lambda$ \cite{blo}, yields \cite{RF-BsKK}
\begin{equation}\label{Bd-ampl2}
A(B_d^0\to\pi^+\pi^-)=e^{i\gamma}\left(1-\frac{\lambda^2}{2}\right){\cal C}
\left[1-d\,e^{i\theta}e^{-i\gamma}\right],
\end{equation}
where
\begin{equation}\label{Aap-def}
{\cal C}\equiv\lambda^3A\,R_b\left(A_{\rm cc}^{u}+A_{\rm pen}^{ut}\right),
\end{equation}
with $A_{\rm pen}^{ut}\equiv A_{\rm pen}^{u}-A_{\rm pen}^{t}$, and
\begin{equation}\label{ap-def}
d\,e^{i\theta}\equiv\frac{1}{(1-\lambda^2/2)R_b}
\left(\frac{A_{\rm pen}^{ct}}{A_{\rm cc}^{u}+A_{\rm pen}^{ut}}\right).
\end{equation}
The quantity $A_{\rm pen}^{ct}$ is defined in analogy to $A_{\rm pen}^{ut}$,
and the CKM factors are given as usual by $\lambda\equiv|V_{us}|=0.22$, 
$A\equiv|V_{cb}|/\lambda^2=0.81\pm0.06$ and 
$R_b\equiv|V_{ub}/(\lambda V_{cb})|=0.41\pm0.07$. The ``penguin parameter'' 
$d\,e^{i\theta}$, which measures -- sloppily speaking -- the ratio of the 
$B_d\to\pi^+\pi^-$ ``penguin'' to ``tree'' contributions, will play a 
central role in this paper. 

Using the Standard-Model parametrization (\ref{Bd-ampl2}), we obtain 
\cite{RF-BsKK}
\begin{eqnarray}
{\cal A}_{\rm CP}^{\rm dir}(B_d\to\pi^+\pi^-)&=&
-\left[\frac{2\,d\sin\theta\sin\gamma}{1-
2\,d\cos\theta\cos\gamma+d^2}\right]\label{ACP-dir}\\
{\cal A}_{\rm CP}^{\rm mix}(B_d\to\pi^+\pi^-)&=&+\left[\,
\frac{\sin(\phi_d+2\gamma)-2\,d\,\cos\theta\,\sin(\phi_d+\gamma)+
d^2\sin\phi_d}{1-2\,d\cos\theta\cos\gamma+d^2}\,\right],\label{ACP-mix}
\end{eqnarray}
where $\phi_d=2\beta$ can be determined with the help of the ``gold-plated'' 
mode $B_d\to J/\psi K_{\rm S}$ through
\begin{equation}\label{Phid-det}
{\cal A}_{\rm CP}^{\rm mix}(B_d\to J/\psi K_{\rm S})=-\sin\phi_d.
\end{equation}
Strictly speaking, mixing-induced CP violation in $B_d\to J/\psi\, K_{\rm S}$ 
probes $\phi_d+\phi_K$, where $\phi_K$ is related to the weak 
$K^0$--$\overline{K^0}$ mixing phase and is negligibly small in the Standard 
Model. Due to the small value of the CP-violating parameter $\varepsilon_K$ 
of the neutral kaon system, $\phi_K$ can only be affected by very contrived 
models of new physics \cite{nirsil}.

In the case of $B_s\to K^+K^-$, we have \cite{RF-BsKK}
\begin{equation}\label{Bs-ampl}
A(B_s^0\to K^+K^-)=e^{i\gamma}\lambda\,{\cal C}'\left[1+\left(
\frac{1-\lambda^2}{\lambda^2}\right)d'e^{i\theta'}e^{-i\gamma}\right],
\end{equation}
where
\begin{equation}
{\cal C}'\equiv\lambda^3A\,R_b\left(A_{\rm cc}^{u'}+A_{\rm pen}^{ut'}\right)
\end{equation}
and 
\begin{equation}\label{dp-def}
d'e^{i\theta'}\equiv\frac{1}{(1-\lambda^2/2)R_b}
\left(\frac{A_{\rm pen}^{ct'}}{A_{\rm cc}^{u'}+A_{\rm pen}^{ut'}}\right)
\end{equation}
correspond to (\ref{Aap-def}) and (\ref{ap-def}), respectively. The primes
remind us that we are dealing with a $\bar b\to\bar s$ transition. It
should be emphasized that (\ref{Bd-ampl2}) and (\ref{Bs-ampl}) are completely
general parametrizations of the $B_d^0\to\pi^+\pi^-$ and $B_s^0\to K^+K^-$ 
decay amplitudes within the Standard Model, relying only on the unitarity 
of the CKM matrix. In particular, these expressions take into account also 
final-state-interaction effects, which received a lot of attention in the
recent literature \cite{FSI}.

Since the decays $B_d\to\pi^+\pi^-$ and $B_s\to K^+K^-$ are related to
each other by interchanging all down and strange quarks, the $U$-spin 
flavour symmetry of strong interactions implies
\begin{equation}\label{U-spin1}
d\,e^{i\theta}=d'e^{i\theta'}.
\end{equation}
Interestingly, this relation is not affected by $U$-spin-breaking corrections 
within a certain model-dependent approach (a modernized version of the 
``Bander--Silverman--Soni mechanism'' \cite{bss}), making use -- among other 
things -- of the ``factorization'' hypothesis to estimate the relevant 
hadronic matrix elements \cite{RF-BsKK}. It would be interesting to
investigate the $U$-spin-breaking corrections to (\ref{U-spin1}) also within
the ``QCD factorization'' approach, which was recently proposed in 
Ref.~\cite{BBNS}. In this paper, it was argued that there is a heavy-quark
expansion for non-leptonic $B$-decays into two light mesons, and that
non-factorizable corrections, as well as final-state-interaction processes,
are suppressed by $\Lambda_{\rm QCD}/m_b$. We shall come back to this 
approach in Section~\ref{sec:d-A-const}, where a comparison of its prediction
for the penguin parameter $d\,e^{i\theta}$ is made with the constraints 
that are implied by the CLEO results (\ref{CLEO-res1}) and (\ref{CLEO-res2}).

For the following considerations, it is useful to introduce the observable
\begin{equation}\label{H-def}
H\equiv\frac{1}{\epsilon}\,\left|\frac{{\cal C}'}{{\cal C}}\right|^2
\left[\frac{M_{B_d}}{M_{B_s}}\,\frac{\Phi(M_K/M_{B_s},M_K/M_{B_s})}{
\Phi(M_\pi/M_{B_d},M_\pi/M_{B_d})}\,\frac{\tau_{B_s}}{\tau_{B_d}}\right]
\left[\frac{\mbox{BR}(B_d\to\pi^+\pi^-)}{\mbox{BR}(B_s\to K^+K^-)}\right],
\end{equation}
where 
\begin{equation}
\epsilon\equiv\frac{\lambda^2}{1-\lambda^2},
\end{equation}
and
\begin{equation}
\Phi(x,y)\equiv\sqrt{\left[1-(x+y)^2\right]\left[1-(x-y)^2\right]}
\end{equation}
denotes the usual two-body phase-space function. The CP-averaged branching 
ratio BR$(B_s\to K^+K^-)$ can be extracted from the corresponding 
``untagged'' rate \cite{RF-BsKK}, where no rapid oscillatory $\Delta M_st$ 
terms are present \cite{dunietz-prd}. In the strict $U$-spin limit, we 
have $|{\cal C}'|=|{\cal C}|$. Corrections to this relation can be calculated 
within the ``factorization'' approximation, yielding
\begin{equation}\label{U-fact}
\left|\frac{{\cal C}'}{{\cal C}}\right|_{\rm fact}=\,
\frac{f_K}{f_\pi}\frac{F_{B_sK}(M_K^2;0^+)}{F_{B_d\pi}(M_\pi^2;0^+)}
\left(\frac{M_{B_s}^2-M_K^2}{M_{B_d}^2-M_\pi^2}\right),
\end{equation}
where $f_K$ and $f_\pi$ denote the kaon and pion decay constants, and the 
form factors $F_{B_sK}(M_K^2;0^+)$ and $F_{B_d\pi}(M_\pi^2;0^+)$ 
parametrize the hadronic quark-current matrix elements 
$\langle K^-|(\bar b u)_{\rm V-A}|B^0_s\rangle$ and 
$\langle\pi^-|(\bar b u)_{\rm V-A}|B^0_d\rangle$, respectively \cite{BSW}.
If we employ (\ref{Bd-ampl2}) and (\ref{Bs-ampl}), we obtain the 
expression
\begin{equation}\label{H-expr}
H=\frac{1-2\,d\,\cos\theta\cos\gamma+d^2}{\epsilon^2+
2\,\epsilon\,d'\cos\theta'\cos\gamma+d'^2},
\end{equation} 
which will play a key role in the following sections. Let us also note 
that there is an interesting relation between $H$ and the corresponding 
direct CP asymmetries \cite{RF-BsKK}:
\begin{equation}\label{CP-rel}
{\cal A}_{\rm CP}^{\rm dir}(B_s\to K^+K^-)=
-\,\epsilon\, H \left(\frac{d'\sin\theta'}{d\,\sin\theta}\right)
{\cal A}_{\rm CP}^{\rm dir}(B_d\to\pi^+\pi^-).
\end{equation}  

Since the decays $B_s\to K^+K^-$ and $B_d\to\pi^\mp K^\pm$ differ only 
in their spectator quarks, we have
\begin{equation}\label{ACP-rep}
{\cal A}_{\rm CP}^{\rm dir}(B_s\to K^+K^-)\approx{\cal A}_{\rm CP}^{\rm dir}
(B_d\to\pi^\mp K^\pm)
\end{equation}
\begin{equation}\label{BR-rep}
\mbox{BR}(B_s\to K^+K^-)
\approx\mbox{BR}(B_d\to\pi^\mp K^\pm)\,\frac{\tau_{B_s}}{\tau_{B_d}},
\end{equation}
and obtain
\begin{equation}\label{H-res}
H\approx\frac{1}{\epsilon}\left(\frac{f_K}{f_\pi}\right)^2
\left[\frac{\mbox{BR}(B_d\to\pi^+\pi^-)}{\mbox{BR}(B_d\to\pi^\mp K^\pm)}
\right]=7.4\pm3.0.
\end{equation}
Here we have also taken into account the CLEO results (\ref{CLEO-res1}) 
and (\ref{CLEO-res2}), and have added the experimental errors in quadrature. 
The advantage of (\ref{H-res}) is that it allows the determination of $H$
without a measurement of the decay $B_s\to K^+K^-$. However, it should be 
kept in mind that this relation relies not only on $SU(3)$ flavour-symmetry 
arguments, but also on a certain dynamical assumption. The point is that 
$B_s\to K^+K^-$ receives also contributions from ``exchange'' and ``penguin 
annihilation'' topologies, which are absent in $B_d\to\pi^\mp K^\pm$. It 
is usually assumed that these contributions play a minor role \cite{ghlr}. 
However, they may be enhanced through certain rescattering effects \cite{FSI}.
Although these topologies do {\it not} lead to any problems in the strategies
discussed below if $H$ is fixed through a measurement of $B_s\to K^+K^-$  -- 
even if they should turn out to be sizeable -- they may affect 
(\ref{ACP-rep})--(\ref{H-res}). The importance of the ``exchange'' and 
``penguin annihilation'' topologies contributing to $B_s\to K^+K^-$ can 
be probed -- in addition to (\ref{ACP-rep}) and (\ref{BR-rep}) -- with the 
help of the decay $B_s\to\pi^+\pi^-$. The na\"\i ve expectation for the 
corresponding branching ratio is ${\cal O}(10^{-8})$; a significant 
enhancement would signal that the ``exchange'' and ``penguin annihilation'' 
topologies cannot be neglected. Another interesting decay in this respect 
is $B_d\to K^+K^-$, for which already stronger experimental constraints 
exist \cite{groro}.

\section{Constraining the Penguin Parameters and the\\  
Direct CP Asymmetries}\label{sec:d-A-const}
If we make use of (\ref{H-expr}) and apply the $U$-spin relation 
(\ref{U-spin1}), the observable $H$ allows us to determine the quantity
\begin{equation}\label{C-def}
C\equiv\cos\theta\,\cos\gamma
\end{equation}
as a function of $d$:
\begin{equation}\label{C-expr}
C=\frac{a-d^2}{2\,b\,d},
\end{equation}
where
\begin{equation}\label{a-b-def}
a=\frac{1-\epsilon^2H}{H-1}\qquad\mbox{and}\qquad b=\frac{1+\epsilon H}{H-1}\,.
\end{equation}
In Ref.~\cite{FM}, a similar function of strong and weak phases was 
considered for the $B_d\to\pi^\mp K^\pm$, $B^\pm\to\pi^\pm K$ system, 
and it was pointed out that this quantity plays an important role to derive 
interesting constraints. Since $C$ is the product of two cosines, it has to 
lie between $-1$ and $+1$, thereby implying an allowed range for $d$. If we 
take into account (\ref{C-expr}) and (\ref{a-b-def}), we obtain (for
$H<1/\epsilon^2$) 
\begin{equation}\label{d-bounds1}
\frac{1-\epsilon\sqrt{H}}{1+\sqrt{H}}\leq d\leq\frac{1+\epsilon\sqrt{H}}{|1-
\sqrt{H}|}.
\end{equation}
An alternative derivation of this range can be found in Ref.~\cite{pirjol}. 
In the special case of $H=1$, there is only a lower bound on $d$, which
is given by $d_{\rm min}=(1-\epsilon)/2$; for $H<1$, $C$ takes a minimal 
value that implies an allowed range for $\gamma$: 
\begin{equation}
|\cos\gamma|\geq C_{\rm min}=\frac{\sqrt{(1-\epsilon^2H)(1-H)}}{1+\epsilon H}
\approx\sqrt{1-H}.
\end{equation}
From a conceptual point of view, this bound on $\gamma$ is completely 
analogous to the one derived in \cite{FM}. Unfortunately, it is only of 
academic interest in the present case, as (\ref{H-res}) indicates $H>1$, 
which we shall assume in the following discussion. So far, we have treated
$\theta$ and $\gamma$ as ``unknown'', free parameters. However, for 
a given value of $\gamma$, we have 
\begin{equation}\label{C-range}
-|\cos\gamma|\leq C\leq+|\cos\gamma|,
\end{equation}
and obtain constraints on $d$ that are stronger than (\ref{d-bounds1}):
\begin{equation}\label{d-range}
d_{\rm min}^{\rm max}=\pm b|\cos\gamma|+\sqrt{a+b^2\cos^2\gamma}.
\end{equation}

\begin{figure}
\centerline{\rotate[r]{
\epsfysize=10.2truecm
{\epsffile{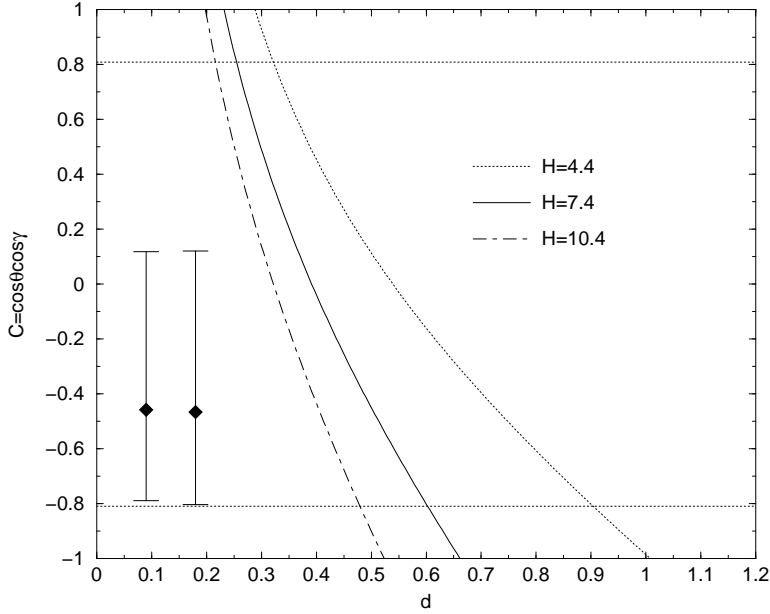}}}}
\caption{The dependence of $C=\cos\theta\cos\gamma$ on the penguin 
parameter $d$ for various values of the observable $H$. The ``diamonds'' 
with the error bars represent the results of the ``QCD factorization'' 
approach \cite{BBNS} for the presently allowed range of $\gamma$, as 
explained in the text. The horizontal dotted lines correspond to 
$C=\pm\cos36^\circ$.}\label{fig:d}
\end{figure}

\begin{figure}
\centerline{\rotate[r]{
\epsfysize=10.2truecm
{\epsffile{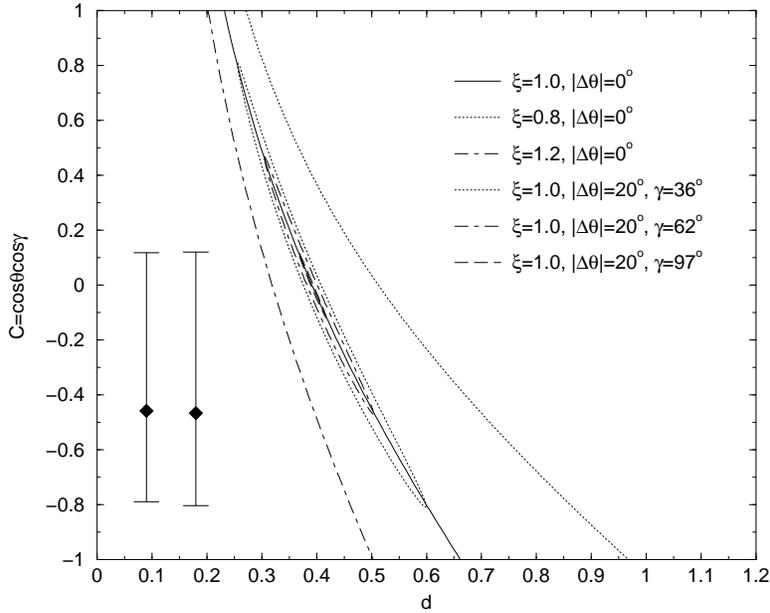}}}}
\caption{The impact of corrections to (\ref{U-spin1}), parametrized 
through $d'=\xi\, d$ and $\theta'=\theta+\Delta\theta$, on the contour
in the $d$--$C$ plane corresponding to $H=7.4$.}\label{fig:break}
\end{figure}

In Fig.~\ref{fig:d}, we show the dependence of $C$ on $d$ for the values of 
the observable $H$ given in (\ref{H-res}). Interestingly, the large values 
of $H$ imply a rather restricted range for $d$. In particular, we get the
lower bound $d\geq0.2$. The ``diamonds'' in Fig.~\ref{fig:d} represent
the results obtained within the ``QCD factorization'' approach \cite{BBNS}, 
representing the state-of-the-art technology in the calculation of the
penguin parameter $d\,e^{i\theta}$:
\begin{equation}\label{facto-res}
\left.d\,e^{i\theta}\right|_{\rm QCD-fact}=0.09\,[0.18]\,e^{i \,193\, 
[187]^\circ}.
\end{equation}
Here a certain formally power-suppressed contribution, which is ``chirally
enhanced'' through the factor
\begin{equation}
r_\chi=\frac{2M_\pi^2}{(m_u+m_d)\,m_b},
\end{equation}
has been neglected [included at leading order]. The ``error bars''
in Fig.~\ref{fig:d} correspond to the presently allowed range for $\gamma$ 
that is implied by the usual ``indirect'' fits of the unitarity 
triangle~\cite{UT-fits}:
\begin{equation}\label{gamma-range}
36^\circ\leq\gamma\leq97^\circ,
\end{equation}
and the ``diamonds'' are evaluated with (\ref{facto-res}) for the
preferred (central) value of $\gamma=62^\circ$. The horizontal dotted lines 
in Fig.~\ref{fig:d} represent $C=\pm\cos36^\circ$. It is an interesting
feature of the contours in the $d$--$C$ plane that they allow in principle
the determination of $\cos\gamma$ with the help (\ref{facto-res}), i.e.\ if
$d$ and $\theta$ are known. However, as can be seen in Fig.~\ref{fig:d},
the most recent CLEO data on $B_d\to\pi^+\pi^-$ and $B_d\to\pi^\mp K^\pm$ 
are not in favour of an interpretation of the ``QCD factorization'' result 
(\ref{facto-res}) within the Standard Model; a solution could be obtained 
for $d\approx0.2$ and $C\approx1$. However, since (\ref{facto-res}) gives 
$\cos\theta\approx-1$, we would then conclude that $\cos\gamma\approx-1$, 
which would be in conflict with the Standard-Model range (\ref{gamma-range}).  
Arguments for $\cos\gamma<0$ using $B\to PP$, $PV$ and $VV$ decays were also 
given in Ref.~\cite{HY}.

Before we discuss the origin of a possible discrepancy of the ``QCD 
factorization'' results with the contours in the $d$--$C$ plane, let 
us have a closer look at the impact of corrections to (\ref{U-spin1}). 
To this end, we generalize this relation as follows:
\begin{equation}\label{break-rel}
d'=\xi\, d,\quad \theta'=\theta+\Delta\theta,
\end{equation}
yielding
\begin{equation}\label{C-gen}
C\equiv\cos\theta\,\cos\gamma=
\left(\frac{1}{1+u^2}\right)\left[\,\frac{a-d^2}{2\,b\,d}\pm 
u\,\sqrt{(1+u^2)\cos^2\gamma-\left(\frac{a-d^2}{2\,b\,d}\right)^2}\,\right],
\end{equation}
where $a$ and $b$ correspond to the following generalization of 
(\ref{a-b-def}):
\begin{equation}\label{a-b-gen}
a=\frac{1-\epsilon^2H}{\xi^2H-1},\quad b=\frac{1+
\epsilon\,\xi H \cos\Delta\theta}{\xi^2H-1},
\end{equation} 
and 
\begin{equation}\label{u-def}
u=\frac{\epsilon\,\xi H\sin\Delta\theta}{1+
\epsilon\,\xi H\cos\Delta\theta}.
\end{equation}
Since the parameter $u$ is doubly suppressed by $\epsilon$ and 
$\Delta\theta$, it is a small quantity. In the case of 
$\Delta\theta=20^\circ$, $\xi=1$ and $H=7.4$, we have, for example, 
$u=0.10$. In Fig.~\ref{fig:break}, we 
illustrate the impact of $\xi\not=1$ and $\Delta\theta\not=0$ on the contour 
in the $d$--$C$ plane corresponding to $H=7.4$. In contrast to (\ref{C-expr}),
the general expression (\ref{C-gen}) depends also on the CKM angle $\gamma$ 
for $\Delta\theta\not=0$. However, since the major effect in 
Fig.~\ref{fig:break} is due to possible corrections to $d'=d$, we shall 
assume $\theta'=\theta$ in the remainder of this paper. In this case, 
(\ref{C-gen}) takes the same form as (\ref{C-expr}).

Although it is too early to draw any definite conclusions, let us note that
there would be basically two different explanations for a discrepancy
of the ``QCD factorization'' results with the contours shown in 
Figs.~\ref{fig:d} and \ref{fig:break}: hadronic effects or physics beyond 
the Standard Model. Concerning the former case, the $\Lambda_{\rm QCD}/m_b$ 
terms and the ``chirally enhanced'' contributions may actually play an 
important role. Interestingly, the inclusion of the latter ones at leading 
order shifts the value of $d$ in the right direction. In order to get
the full picture, it would be an important task to analyse (\ref{U-fact}) 
and (\ref{break-rel}) in the ``QCD factorization'' approach. Using
present data, it seems that the ``QCD factorization'' results (\ref{facto-res})
can only be accommodated -- if at all possible -- for values of $\gamma$ 
sizeably larger than $90^\circ$, which would be in conflict with
(\ref{gamma-range}), and a possible sign for new physics. Since the parameter 
$d\,e^{i\theta}$ is governed by penguin topologies, i.e.\ by 
flavour-changing neutral-current (FCNC) processes, it may well be affected 
by physics beyond the Standard Model \cite{new-phys,Fle-Mat}. Moreover, 
it should be kept in mind that the unitarity of the CKM matrix has been used 
in the calculation of the contours shown in Figs.~\ref{fig:d} and 
\ref{fig:break}. Further studies and better data are needed to explore these 
exciting issues in more detail.

Let us now turn to the constraints on the direct CP asymmetries (see also
\cite{pirjol,FM}). Before turning to the general case, it is instructive
to consider $\gamma=90^\circ$. In this case, we obtain
\begin{equation}\label{ACP90}
\left.{\cal A}_{\rm CP}^{\rm dir}(B_d\to\pi^+\pi^-)\right|_{\gamma=90^\circ}=
-\left[\frac{2\,d\sin\theta}{1+d^2}\right],\quad
\left.{\cal A}_{\rm CP}^{\rm dir}(B_s\to K^+K^-)\right|_{\gamma=90^\circ}=
+\left[\frac{2\,\epsilon\,d'\sin\theta'}{\epsilon^2+d'^2}\right],
\end{equation}
and
\begin{equation}\label{H90}
\left.H\right|_{\gamma=90^\circ}=\frac{1+d^2}{\epsilon^2+d'^2}.
\end{equation}
The CP asymmetries given in (\ref{ACP90}) take their extremal values for 
$\theta=\theta'=\pm90^\circ$, and (\ref{H90}) allows us to determine $d$:
\begin{equation}
\left.d\right|_{\gamma=90^\circ}=\sqrt{\frac{1-\epsilon^2H}{\xi^2H-1}},
\end{equation}
where we have also used $d'=\xi\,d$. Consequently, we obtain
\begin{equation}\label{ACPBdpipi-approx}
\left|{\cal A}_{\rm CP}^{\rm dir}(B_d\to\pi^+\pi^-)
\right|_{\gamma=90^\circ}^{\rm max}=
2\,\sqrt{\frac{(1-\epsilon^2H)(\xi^2H-1)}{\left(\xi^2-\epsilon^2
\right)^2H^2}}\approx\frac{2}{\xi\sqrt{H}}
\end{equation}
and
\begin{equation}\label{ACPBsKK-approx}
\left|{\cal A}_{\rm CP}^{\rm dir}(B_s\to K^+ K^-)
\right|_{\gamma=90^\circ}^{\rm max}=2\,\epsilon\,\xi\,\sqrt{\frac{(1-
\epsilon^2H)(\xi^2H-1)}{\left(\xi^2-\epsilon^2\right)^2}}
\approx2\,\epsilon\,\sqrt{H}.
\end{equation}
Let us emphasize that (\ref{ACPBsKK-approx}) is essentially {\it unaffected}
by any corrections to the $U$-spin relation (\ref{U-spin1}) for 
$H={\cal O}(10)$; its theoretical accuracy is practically only limited by 
(\ref{U-fact}), which enters in the determination of $H$ through (\ref{H-def}).

\begin{figure}
\centerline{\rotate[r]{
\epsfysize=10.2truecm
{\epsffile{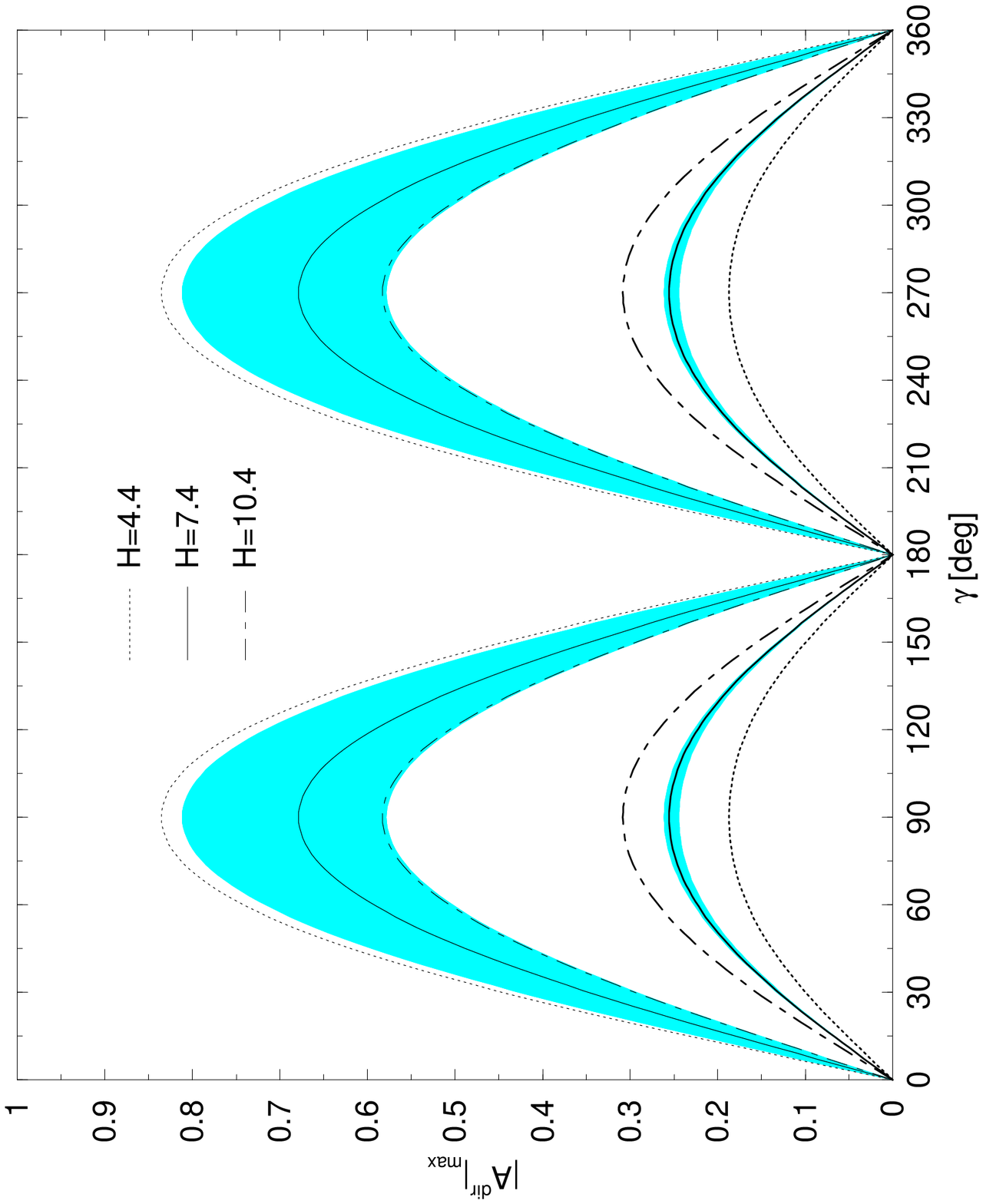}}}}
\caption{The dependence of the maximally allowed direct CP-violating 
asymmetries 
$\left.|{\cal A}_{\rm CP}^{\rm dir}(B_d\to\pi^+\pi^-)\right|_{\rm max}$ 
(thin lines) and $\left.|{\cal A}_{\rm CP}^{\rm dir}(B_s\to K^+K^-)
\right|_{\rm max}\approx\left.|{\cal A}_{\rm CP}^{\rm dir}(B_d\to\pi^\mp K^\pm)
\right|_{\rm max}$ (thick lines) on the CKM angle $\gamma$ for various values 
of the observable $H$. The shaded \mbox{regions} correspond to a variation 
of $\xi$ within the interval $[0.8,1.2]$ for $H=7.4$.}\label{fig:Adir}
\end{figure}

In the general case $\gamma\not=90^\circ$, we employ (\ref{C-expr})
to eliminate the CP-conserving strong phase $\theta$ in (\ref{ACP-dir}). 
Following these lines, we obtain 
${\cal A}_{\rm CP}^{\rm dir}(B_d\to\pi^+\pi^-)$ as a function of $d$ for
a given value of $\gamma$. If we keep $\gamma$ fixed, and vary $d$ within 
the allowed range corresponding to (\ref{d-range}), we find that 
$|{\cal A}_{\rm CP}^{\rm dir}(B_d\to\pi^+\pi^-)|$ takes the following 
maximal value:
\begin{equation}\label{ACPmax1}
\left.|{\cal A}_{\rm CP}^{\rm dir}(B_d\to\pi^+\pi^-)\right|_{\rm max}=
2|\sin\gamma|\sqrt{\frac{a+b^2\cos^2\gamma}{(1+a)^2-4(a-b)(1+b)\cos^2\gamma}},
\end{equation}
where $a$ and $b$ are given in (\ref{a-b-gen}) for $\Delta\theta=0$ (see
the comment after (\ref{u-def})). In the case of $B_s\to K^+K^-$, we obtain
\begin{eqnarray}
\lefteqn{\left.|{\cal A}_{\rm CP}^{\rm dir}(B_d\to\pi^\mp K^\pm)
\right|_{\rm max}\approx\left.|{\cal A}_{\rm CP}^{\rm dir}(B_s\to K^+K^-)
\right|_{\rm max}}\nonumber\\
&&\qquad=2\,\epsilon\,\xi\,H\,|\sin\gamma|
\sqrt{\frac{a+b^2\cos^2\gamma}{(1+a)^2-4(a-b)(1+b)
\cos^2\gamma}}.\label{ACPmax2}
\end{eqnarray}

For $\gamma=90^\circ$, these expressions reduce to (\ref{ACPBdpipi-approx}) 
and (\ref{ACPBsKK-approx}), respectively. In Fig.~\ref{fig:Adir}, we show 
the dependence of (\ref{ACPmax1}) and (\ref{ACPmax2}) on $\gamma$ for the 
values of $H$ given in (\ref{H-res}). The shaded regions correspond to a 
variation of the 
parameter $\xi\equiv d'/d$ within the interval $[0.8,1.2]$ for $H=7.4$. In 
contrast to (\ref{ACPmax1}), (\ref{ACPmax2}) is essentially unaffected by a 
variation of $\xi$, as we have already noted above. The range for $H$ given 
in (\ref{H-res}) disfavours large direct CP violation in $B_s\to K^+K^-$ and 
$B_d\to\pi^\mp K^\pm$ (see also \cite{Fle-Mat}), which is also consistent 
with the 90\% C.L.\ interval of 
$-0.22\leq {\cal A}_{\rm CP}^{\rm dir}(B_d\to\pi^\mp K^\pm)\leq +0.30$ 
reported recently by the CLEO collaboration~\cite{CLEO2}. On the other 
hand, there is a lot of space for large direct CP violation in 
$B_d\to\pi^+\pi^-$. As can be seen in Fig.~\ref{fig:Adir}, a measurement 
of non-vanishing CP asymmetries $|{\cal A}_{\rm CP}^{\rm dir}|_{\rm exp}$ 
would allow us to exclude immediately a certain range of $\gamma$ around 
$0^\circ$ and $180^\circ$, as values of $\gamma$ corresponding to 
$|{\cal A}_{\rm CP}^{\rm dir}|_{\rm exp}>
|{\cal A}_{\rm CP}^{\rm dir}|_{\rm max}$ are excluded. However, in 
order to constrain this CKM angle, the mixing-induced CP asymmetry 
${\cal A}_{\rm CP}^{\rm mix}(B_d\to\pi^+\pi^-)$ appears to be more 
powerful. 

Before we turn to these bounds in the following section, let us note 
that the observables of the decay $B_d\to\pi^+\pi^-$ were combined with
the CP-averaged $B_d\to\pi^\mp K^\pm$ and $B_s\to K^+K^-$ branching 
ratios in Refs.~\cite{charles} and \cite{pirjol}, respectively, to 
derive constraints on the penguin effects in the extraction of the
CKM angle $\alpha$. In the present paper, we combine the experimental 
information provided by these modes in a different way, which appears
more favourable to us. In particular, we use the mixing-induced CP 
asymmetry of the ``gold-plated'' mode $B_d\to J/\psi K_{\rm S}$ as an 
additional input \cite{RF-BsKK}, and derive bounds on the CKM angle
$\gamma$. The utility of $B_d\to\pi^\mp K^\pm$ decays to control the
penguin effects on CP violation in  $B_d\to\pi^+\pi^-$ was also
emphasized in Ref.~\cite{SW}.

\boldmath
\section{Constraining the CKM Angle $\gamma$}\label{sec:gamma-const}
\unboldmath
In the following discussion, we assume that $\phi_d=2\beta$ has been 
measured at the $B$-factories through (\ref{Phid-det}), which is one of
the major goals of these experiments. The presently allowed range for 
$\beta$ that is implied by the usual ``indirect'' fits of the unitarity 
triangle is given as follows~\cite{UT-fits}:
\begin{equation}\label{beta-range}
16^\circ\leq\beta\leq35^\circ,
\end{equation}
with a preferred (central) value of $\beta=25^\circ$, which is also 
consistent with the present experimental result ${\cal A}_{\rm CP}^{\rm mix}
(B_d\to J/\psi K_{\rm S})=-\sin(2\beta)=-0.79^{+0.44}_{-0.41}$ of the CDF 
collaboration \cite{CDF}. A measurement of this mixing-induced CP asymmetry 
allows us to determine only $\sin\phi_d$, i.e.\ to fix $\phi_d$ up to a 
twofold ambiguity. Several strategies were proposed in the literature to 
resolve this ambiguity \cite{ambig}. In the $B$-factory era, an experimental 
uncertainty of $\left.\Delta\sin\phi_d\right.|_{\rm exp}=0.05$ seems to be 
achievable after a few years of taking data, which corresponds to an
uncertainty of $\Delta\phi_d=\pm5^\circ$ for the central value of 
$\phi_d=50^\circ$.

If we assume, for a moment, that there are no penguin effects present in 
$B_d\to\pi^+\pi^-$, i.e.\ $d=0$, we would simply have 
\begin{equation}\label{no-pen}
\left.{\cal A}_{\rm CP}^{\rm mix}(B_d\to\pi^+\pi^-)\right|_{d=0}=
\sin(\phi_d+2\gamma),
\end{equation}
as can be seen in (\ref{ACP-mix}). Since the unitarity of the CKM matrix 
implies $\phi_d+2\gamma=-2\alpha$, this CP asymmetry is usually written 
as ${\cal A}_{\rm CP}^{\rm mix}(B_d\to\pi^+\pi^-)|_{d=0}=-\sin(2\alpha)$, 
and would allow a direct measuerment of $\alpha$. However, (\ref{no-pen}) 
is the ``generic'' interpretation of this CP asymmetry, allowing us to 
determine $\gamma$, if $\phi_d$ is fixed through $B_d\to J/\psi K_{\rm S}$. 
In the case of large penguin contributions, this interpretation of 
${\cal A}_{\rm CP}^{\rm mix}(B_d\to\pi^+\pi^-)$ actually appears to be more 
favourable than the usual one in terms of $\alpha$, which was employed, 
for example, in Refs.~\cite{charles,pirjol}. Since we definitely have to 
worry about penguin effects in $B_d\to\pi^+\pi^-$, as we have pointed out 
in the previous section, we shall use the corresponding mixing-induced CP 
asymmetry to contrain the CKM angle $\gamma$ in this section.

Concerning the search for new physics, $\gamma$ is actually the interesting
aspect of the mixing-induced $B_d\to\pi^+\pi^-$ CP asymmetry. If $\phi_d$ 
is affected by new physics, these effects could be seen, for example, by 
comparing the $B_d\to J/\psi K_{\rm S}$ results with the ``indirect'' range 
(\ref{beta-range}). Since this channel is governed by $\bar b\to\bar cc\bar s$
``tree'' processes, its decay amplitude is not expected to be affected 
significantly by new-physics effects, and allows the determination of 
$\phi_d$ even in the presence of physics beyond the Standard Model. 
In order to search for indications of new physics, the values of $\gamma$ 
implied by the CP-violating effects in $B_d\to\pi^+\pi^-$ could be compared 
with the ``indirect'' range arising from the usual fits of the unitarity 
triangle, or with theoretically clean extractions from pure ``tree'' decays, 
such as $B_d\to D^{\ast\pm}\pi^\mp$ or $B\to DK$ (see also the brief 
discussion of new-physics effects in Section~\ref{sec:d-A-const}).

\begin{figure}
\centerline{\rotate[r]{
\epsfysize=10.2truecm
{\epsffile{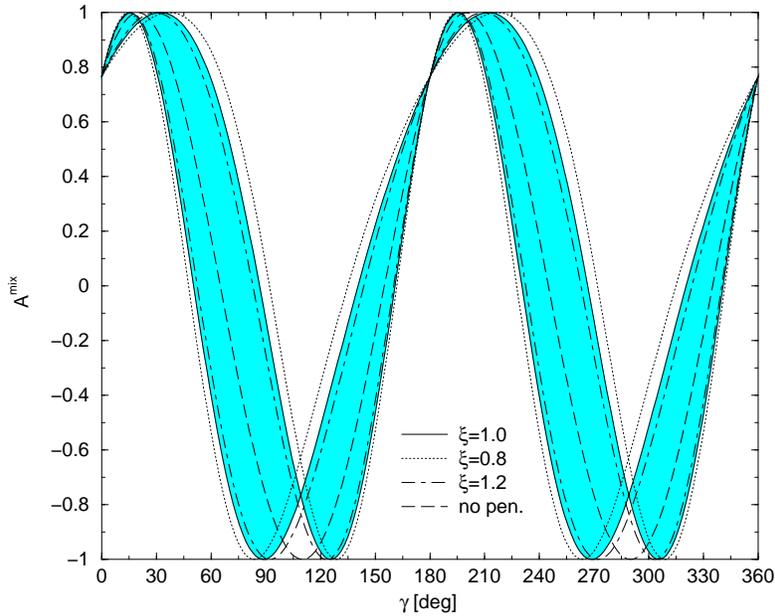}}}}
\caption{The dependence of the allowed range for 
${\cal A}_{\rm CP}^{\rm mix}(B_d\to\pi^+\pi^-)$ on the CKM angle $\gamma$
for $H=7.4$ and $\phi_d=50^\circ$.}\label{fig:Amix1d}
\end{figure}

If we look at the expressions (\ref{ACP-dir}) and (\ref{ACP-mix}) for
the direct and mixing-induced CP asymmetries of the decay $B_d\to\pi^+\pi^-$,
we observe that the CP-conserving strong phase $\theta$ enters only 
in the form of $\cos\theta$ in the latter case. Consequently, using 
$\cos\theta=C/\cos\gamma$ and (\ref{C-expr}), we obtain 
\begin{equation}\label{ACPmix-funct}
{\cal A}_{\rm CP}^{\rm mix}(B_d\to\pi^+\pi^-)
=\frac{\left[\,b\,\sin(\phi_d+2\gamma)\cos\gamma-a\,\sin(\phi_d+\gamma)
\right]+\left[\,\sin(\phi_d+\gamma)+b\,\sin\phi_d\cos\gamma
\right]d^2}{\left[(b-a)+(1+b)\,d^2\right]\cos\gamma},
\end{equation}
where $a$ and $b$ are given in (\ref{a-b-gen}) for $\Delta\theta=0$, i.e.
the small corrections due to $\Delta\theta\not=0$ have been neglected for 
simplicity (see the comment after (\ref{u-def})). Since (\ref{ACPmix-funct}) 
is a monotonic function of the variable $d^2$, it takes its extremal values 
for the minimal and maximal values of $d$ given in (\ref{d-range}); 
inserting them into (\ref{ACPmix-funct}) yields
\begin{equation}
\left.{\cal A}_{\rm CP}^{\rm mix}(B_d\to\pi^+\pi^-)\right|_{\rm extr.}=
\frac{\sin(\phi_d+2\gamma)+a\sin\phi_d+2\,w_{\pm}\left[\sin(\phi_d+\gamma)
+b\,\cos\gamma\sin\phi_d\right]}{1+a+2\,w_{\pm}(1+b)\cos\gamma},
\end{equation}
where
\begin{equation}
w_{\pm}=b\cos\gamma\pm\sqrt{a+b^2\cos^2\gamma}.
\end{equation}

In Fig.~\ref{fig:Amix1d}, we illustrate the resulting allowed range for
${\cal A}_{\rm CP}^{\rm mix}(B_d\to\pi^+\pi^-)$ in the case of $H=7.4$
and $\phi_d=50^\circ$ (shaded region). The impact of a deviation
of the parameter $\xi$ from 1 is illustrated by the dotted and 
dot-dashed lines, which correspond to $\xi=0.8$ and $1.2$, respectively.
For a given value of $\gamma$, the allowed range for the mixing-induced 
$B_d\to\pi^+\pi^-$ CP asymmetry is usually very large. However, a measured 
value of ${\cal A}_{\rm CP}^{\rm mix}(B_d\to\pi^+\pi^-)$ 
would, on the other hand, imply a rather restricted range for $\gamma$. If 
we assume, for example, that 
${\cal A}_{\rm CP}^{\rm mix}(B_d\to\pi^+\pi^-)=0.4$ has been measured, 
and take into account that the experimental value of $\varepsilon_K$ 
implies $\gamma\in[0^\circ,180^\circ]$, we would conclude that 
$41^\circ\leq\gamma\leq74^\circ$ or $158^\circ\leq\gamma\leq170^\circ$.
Allowing $\xi\in[0.8,1.2]$, i.e.\ symmetry-breaking corrections of $20\%$,
we would obtain the slightly modified ranges 
$39^\circ\leq\gamma\leq80^\circ$ $\lor$ $155^\circ\leq\gamma\leq170^\circ$
and $43^\circ\leq\gamma\leq71^\circ$ $\lor$ $160^\circ\leq\gamma\leq170^\circ$
for $\xi=0.8$ and 1.2, respectively. Since the allowed region for $d$ is 
enlarged (reduced) for smaller (larger) values of $H$, the bounds on 
$\gamma$ become weaker (stronger) in this case. 

Let us finally note that if in addition to a measurement of $H$ and 
${\cal A}_{\rm CP}^{\rm mix}(B_d\to\pi^+\pi^-)$ direct CP violation in 
$B_d\to\pi^+\pi^-$ or $B_d\to\pi^\mp K^\pm$ is observed, we have three 
independent observables at our disposal, which depend on $\gamma$, $d$ and 
$\theta$. Consequently, we are then not only in a position to constrain 
these ``unknown'' parameters, but also to {\it determine} them 
\cite{RF-BsKK}. Moreover, the normalization $|{\cal C}|$ of the 
$B_d\to\pi^+\pi^-$ decay amplitude (see (\ref{Bd-ampl2})) can be 
extracted from the corresponding CP-averaged branching ratio, and can
be compared with theoretical predictions. The $B_d\to\pi^\mp K^\pm$ decays 
offer also alternative strategies to determine $\gamma$ and certain hadronic 
quantities, if these transitions are combined with other $B\to \pi K$ 
modes \cite{BpiK}.

\section{Conclusions and Outlook}\label{sec:concl}
The decays $B_d\to\pi^+\pi^-$ and $B_s\to K^+K^-$ provide interesting
strategies to extract the CKM angle $\gamma$ and hadronic penguin parameters 
at ``second-generation'' $B$-physics experiments of the LHC era. In 
this paper, we have considered a variant of this approach for the 
``first-generation'' $B$-factories, where the $B_s\to K^+K^-$ decays 
are replaced by $B_d\to\pi^\mp K^\pm$ modes. 

We have pointed out that the CP-averaged $B_d\to\pi^+\pi^-$ and 
$B_d\to\pi^\mp K^\pm$ branching ratios allow us to fix contours in the 
$d$--$[\cos\theta\cos\gamma]$ plane, which can be compared with theoretical 
results for the $B_d\to\pi^+\pi^-$ ``penguin parameter'' $d\,e^{i\theta}$, 
for example with those of the ``QCD factorization'' approach. Although it 
is too early to draw any definite conclusions, it is interesting to note 
that the most recent CLEO data are not in favour of an interpretation of the 
``QCD factorization'' results within the Standard Model. This feature may 
be due to hadronic effects or new physics. Further theoretical studies
and better experimental data are required to investigate these exciting
issues in more detail.

Another interesting aspect of the recent CLEO results for the CP-averaged 
$B_d\to\pi^+\pi^-$ and $B_d\to\pi^\mp K^\pm$ branching ratios is that 
they imply upper bounds on the corresponding direct CP asymmetries, which
are given by $|{\cal A}_{\rm CP}^{\rm dir}(B_d\to\pi^+\pi^-)|_{\rm max}
\mathrel{\hbox{\rlap{\hbox{\lower4pt\hbox{$\sim$}}}\hbox{$<$}}}0.8$
and $|{\cal A}_{\rm CP}^{\rm dir}(B_d\to\pi^\mp K^\pm)|_{\rm max}
\approx|{\cal A}_{\rm CP}^{\rm dir}(B_s\to K^+K^-)|_{\rm max}
\mathrel{\hbox{\rlap{\hbox{\lower4pt\hbox{$\sim$}}}\hbox{$<$}}}0.3$. The
latter bound is remarkably stable under $U$-spin-breaking corrections --
in contrast to the former one -- and may also play an important role to 
search for new physics. 

If in addition to the CP-averaged $B_d\to\pi^+\pi^-$ and 
$B_d\to\pi^\mp K^\pm$ branching ratios mixing-induced CP violation in 
the former decay is measured, and the $B^0_d$--$\overline{B^0_d}$ mixing 
phase is fixed through $B_d\to J/\psi K_{\rm S}$, interesting constraints
on $\gamma$ can be obtained. A further step in this programme would be
the observation of direct CP violation in $B_d\to\pi^+\pi^-$ or 
$B_d\to\pi^\mp K^\pm$, which would allow a determination of $\gamma$, 
$d\,e^{i\theta}$ and $|{\cal C}|$. In this way, two of the major goals of 
the $B$-factories -- time-dependent analyses of the benchmark modes 
$B_d\to J/\psi K_{\rm S}$ and $B_d\to\pi^+\pi^-$ -- can be combined with 
each other to probe the CKM angle $\gamma$ and to obtain valuable insights 
into the world of penguins. 

Another important step would be a measurement of the CP-averaged 
$B_s\to K^+K^-$ branching ratio, which may be possible at HERA-B and Run II 
of the Tevatron. Using this observable, a certain dynamical assumption 
concerning ``exchange'' and ``penguin annihilation'' topologies can be 
avoided, which has to be made in the case of $B_d\to\pi^\mp K^\pm$. The 
theoretical accuracy would then only be limited by $U$-spin-breaking effects 
and would not be affected by any final-state-interaction processes. The 
final goal is a measuerment of the CP-violating observables of 
$B_s\to K^+K^-$, which should be possible at LHCb and BTeV. At these 
experiments, the physics potential of $B_d\to\pi^+\pi^-$ and 
$B_s\to K^+K^-$ can be fully exploited, and in addition to an extraction 
of $\gamma$ at the level of a few degrees, also interesting consistency 
checks of the basic $U$-spin relations can be performed.

\end{document}